\documentclass[apj,numberedappendix]{emulateapj}
\usepackage{graphics,epsf}
\usepackage{amsmath}                
\usepackage{amsfonts}               
\usepackage{amssymb}                
\usepackage{epsfig}                 
\usepackage{graphicx}
\usepackage{float}

\newcommand{\cm}{{~\rm cm}}
\newcommand{\km}{{~\rm km}}
\newcommand{\s}{{~\rm s}}

\newcommand{\g}{{~\rm g}}

\newcommand{\K}{{~\rm K}}
\newcommand{\erg}{{~\rm erg}}
\newcommand{\yr}{{~\rm yr}}

\newcommand{\AU}{{~\rm AU}}

\begin{document}

\title{Pre-supernova outbursts of massive stars in the presence of a neutron star companion}

\author{Barak Danieli\altaffilmark{1}, \& Noam Soker\altaffilmark{1,2}}

\altaffiltext{1}{Department of Physics, Technion -- Israel Institute of Technology, Haifa
32000, Israel; Barakd@campus.technion.ac.il; soker@physics.technion.ac.il}
\altaffiltext{2}{Guangdong Technion Israel Institute of Technology, Shantou 515069, Guangdong Province, China}


\begin{abstract}
We study the pre-explosion outbursts (PEOs) of massive stars that might result from a rapid expansion of the massive star in the presence of a close companion. We assume that activity in the core of the massive star, an initial mass of $15 M_\odot$,  about two years before explosion energizes the envelope, and with the stellar evolutionary code \textsc{mesa} follow the inflated envelope as a result of energy deposition to the envelope. We examine the conditions for a companion star to accrete mass from the inflated envelope. 
We find that for the general conditions that we assume, bright PEOs require a neutron star companion at an orbital separation of $\approx 1000-2000 R_\odot$. We assume that the mass-accreting neutron star launches jets. These jets shape the circumstellar matter to highly non-spherical structures, such that the explosions of core collapse supernovae (CCSNe) that follow PEOs might lack an axial (cylindrical) symmetry. 
In some case main sequence star companions can also energize PEOs, but much weaker ones. 
This study adds another scenario by which neutron stars can power the radiation  of PEOs. Another scenario is the common envelope jets supernova (CEJSN) impostor where a neutron star enters the envelope of the massive star. 
\end{abstract}


\section{INTRODUCTION}
\label{sec:intro}

Some massive stars undergo pre-explosion outbursts (PEOs) tens of years to days before they terminally explode as a core collapse supernova (CCSN; e.g., \citealt{Foleyetal2007, Pastorelloetal2007, Smithetal2010, Mauerhanetal2013, Ofeketal2013, Pastorelloetal2013, Marguttietal2014, Ofeketal2014, SvirskiNakar2014, Fraseretal2015, Moriya2015, Goranskijetal2016, Ofeketal2016, Tartagliaetal2016, BoianGroh2017,  Marguttietal2017, Liuetal2017, Nyholmetal2017, Pastorelloetal2017, Yaronetal2017}).
The outburst is accompanied by mass ejection that forms a dense circumstellar matter (CSM). After explosion the supernova ejecta collides with the CSM, turning kinetic energy to radiation.  Some of the PEOs are observed to be non-spherical. \cite{Reillyetal2017} deduce from their spectropolarimetry observations of the 2012 PEO of SN~2009ip, that was a major outburst of a luminous blue variable (LBV), that the CSM that was formed from the PEO is compatible with a disk-like geometry.
In some cases enhanced mass loss rate episodes might occur as early as the core carbon-burning phase (e.g., \citealt{Moriyaetal2014, Marguttietal2017}).

Since standard stellar evolutionary simulations do not lead to PEOs, researchers have introduced extra mechanisms to trigger and power PEO of CCSNe. 
Some mechanisms attribute the instability to the envelope of the massive star, such as the radiation-driven instabilities (e.g., \citealt{BlaesSocrates2003}) that might occur in some LBVs (e.g.,  \citealt{Kiriakidisetal1993, Kashietal2016}).
Other mechanisms start from the very high power of the nuclear burning in the core that triggers vigorous core convection. Energy that is carried from the convective zones to the envelope, e.g., by waves \citep{QuataertShiode2012, ShiodeQuataert2014}, causes the envelope to either eject mass (e.g., \citealt{QuataertShiode2012, ShiodeQuataert2014}), or to expand \citep{Soker2013, ShiodeQuataert2014, SmithArnett2014}, or both. 
\cite{Fuller2017} find that waves triggered by the core mainly cause envelope expansion, and by themselves unbind only a small amount of envelope mass (\citealt{FullerRo2018}, however, claim that this mechanism can drive mass-loss in some hydrogen-deficient massive stars). 

\cite{SokerGilkis2017b} propose that the vigorous convection amplifies magnetic fields in the core, and magnetic flux tubes that buoy from the core to the envelope carry the energy to the envelope and cause envelope expansion. To trigger a powerful PEO this scenario requires the presence of a close binary companion that accretes mass from the inflated envelope. 
   
 Because PEOs occur in only about 10 per cent of all CCSNe (e.g., \citealt{Bilinskietal2015, Marguttietal2017}), the mechanism for PEOs must include a rare ingredient.
\cite{ShiodeQuataert2014} estimate that about $20 \%$ of the CCSN progenitors might excite outward propagating waves with $10^{46}-10^{48} \erg$ that trigger PEOs. In the scenario of magnetic activity the rare ingredient is a minimum core rotation that is required to form a strong dynamo. If a close binary companion is required for a powerful PEO, the requirement for its presence further reduces the probability for PEOs. 

Whether PEOs require the presence of a binary companion or not is an open question   (e.g., \citealt{Levesqueetal2014, Marguttietal2017}). 
While single-star PEO processes exist (e.g., \citealt{Shaviv2000, Shaviv2001, Owockietal2004, Quataertetal2016, Moriya2014}), in the present study we adopt the view that powerful PEOs require the presence of a  secondary star that orbits the massive star at a close distance and accretes mass from the inflated envelope (e.g., \citealt{KashiSoker2010, Soker2013, McleySoker2014}). We do emphasize that even PEO binary models require that the massive star first experiences some kind of unstable phase that triggers a strong binary interaction.
 
\cite{McleySoker2014} run the stellar evolutionary code \textsc{mesa} and find that energy deposited to the envelope is likely to lead to its expansion rather than to large mass ejection (also \citealt{Fuller2017}). We here continue their study and calculate the properties of the possible binary companion that can accrete mass, and the energy that the companion might release by the accretion process from the inflated envelope. We find that for the specific instability we use for the primary star a very energetic outburst requires the companion to be a neutron star (NS), and hence in this study we consider mainly a NS companion. 

In section \ref{sec:inflated} we discuss PEOs in cases where the supernova progenitors suffers instability and inflates its envelope. We describe the inflation of the envelope because of the energy we inject into it (section \ref{sec:evolution}), we examine the possible mass of the secondary star (section \ref{sec:companion}), we estimate the accretion power of the secondary star during the pre-explosion outburst (section \ref{sec:accretion}), and we discuss the emission from the PEO (section \ref{sec:bump}). In section \ref{sec:CSM} we study the case of an  expanding shell that might be powered in part by a companion. We summarize in section \ref{sec:summary}.

\section{Accretion from an inflated envelope}
\label{sec:inflated}
\subsection{Envelope inflation}
\label{sec:evolution}
   
We run the stellar evolution code \textsc{mesa} (version 10000; \citealt{Paxtonetal2011, Paxtonetal2013, Paxtonetal2015}) and follow the evolution of a star with an initial mass of $M_{1,0}=15 M_\odot$ and metallicity of $Z=0.02$.  Just before core collapse the mass of the star is $M_1=13.55 M_\odot$ and its radius is $R_1=861 R_\odot$. 
Since we are aiming to present the basic characteristics of the scenario, rather than to explain a specific object or conduct a study of the parameter space, we take these initial mass and metallicity because many studies take these values to represent general CCSNe.   

To mimic core activity that powers the envelope, by waves or by magnetic activity, we inject energy to the envelope and follow the evolution of the structure of the inflated envelope. The energy injection scheme is similar, but not identical, to that of \cite{McleySoker2014}. 
About two years before core collapse we start to inject energy with a power of $L_{\rm wave}=3.2 \times 10^{5} L_\odot$, which is much more than the stellar luminosity at that stage, $L=8.1 \times 10^4 L_\odot$. The wave power is taken from table 2 of \cite{ShiodeQuataert2014} for a non-rotating red supergiant model during core neon burning. We inject the energy inside one numerical shell at the driven radius $r_{d}$ where the wave luminosity equals the maximum luminosity that can be carried by convection $L_{\rm wave}=L_{\rm max,conv} (r_{d}) $. 
The maximum energy that can be carried by convection is 
$L_{\rm max,conv} = 4 \pi \rho r^2 c_s^3$, where $c_s$ is the local sound speed and $\rho$ is the density. 
 The driven radius $r_d$ is calculated at each time step. At the beginning of the injection phase the driven radius is $r_d=845 R_\odot$ where the density is $\rho(r_d) = 2.5 \times 10^{-9} \g \cm^{-3}$, while at the end of the injection phase the driven radius is $r_d=1255 R_\odot$ and the density there is $\rho(r_d) = 5 \times 10^{-10} \g \cm^{-3}$. Namely, at the end of the energy-injection phase the shell into which we inject the energy is in the inflated envelope. Along the entire evolution the injection takes place in a convective zone. 
  
\begin{figure}
\begin{center}
\includegraphics[width=0.45\textwidth]{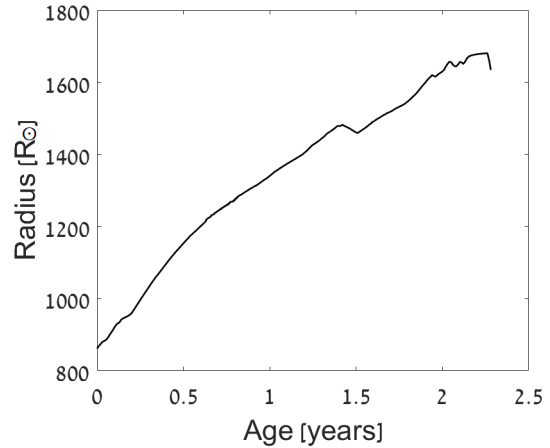}
\caption{The outer radius of the inflated envelope as function of time during the energy injection phase that last for $2.3 \yr$. The star explodes at the end of that phase.   }   
\label{fig:Radius15}
\end{center}
\end{figure}

\begin{figure}
\begin{center}
\includegraphics[width=0.45\textwidth]{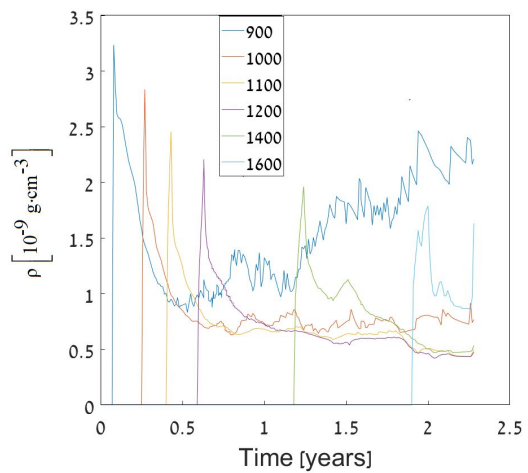}
\caption{
The density as function of time during the envelope inflation phase at several radii, all outside the stellar envelope before inflation starts. Density is in units of $10^{-9} \g \cm^{-3}$. 
}   
\label{fig:Density15}
\end{center}
\end{figure}

We start to inject energy at an age of $1.2704215\times 10^7 \yr$ which we take as $t=0$, and end it at $t=2.3 \yr$ (an age of $1.27042173\times10^7 \yr$). The envelope inflates from an initial radius of $R_1=861 R_\odot$ to a large radius as we present in Fig. \ref{fig:Radius15}.
 In our simulation the star expands without an impulsive mass loss episode. Although \cite{ShiodeQuataert2014} argue that energy deposition as we simulate here leads to an implosive mass loss,  \cite{McleySoker2014} and \cite{Fuller2017} already argued that the main effect of the energy deposition is stellar expansion rather than mass loss.  
In Fig. \ref{fig:Density15} we present the density at six different radii, all were outside the star before inflation started. 
   
\subsection{The companion}
\label{sec:companion}

 The scenario we study in this paper starts with two main sequence massive stars. The more massive star evolves and experiences a CCSN explosion that leaves a NS at an orbital separation of several astronomical units. If at this stage the initially less massive star has a mass of $\ga 9 M_\odot$ the system evolves to a second CCSN. Note that the initial main sequence mass of the less massive star might be below the lower limit for supernova explosion as long as it gains enough mass from the initially more massive star to become a CCSN progenitor, i.e., has a mass of $\ga 9 M_\odot$ after the first CCSN event. 
   
 We study the few years before the second CCSN takes place. During that phase the system composed of a giant star, that we called the primary star, and a secondary star which is a NS. 
The orbital separation is several astronomical units. We assume that the primary giant star experiences an internal instability that deposits energy to its envelope and causes it to expand. In section \ref{sec:evolution} we presented the envelope inflation phase. We now turn to discuss the role of the secondary star. 

To survive outside the envelope of the primary star the secondary star should obey two conditions as follows. (1) Not to be too massive to cause the primary star to overflow its Roche lobe. (2) Be massive enough to maintain Darwin stability against rapid spiraling-in evolution towards the primary envelope. 
 Although we present the results when the first condition is included, it is not necessary at all cases. In cases when Roche lobe overflow (RLOF) takes place and the secondary star accretes mass at a high rate, the secondary star might spiral-in into the envelope of the primary star. However, in many cases RLOF will proceed on a slow rate, and even if the secondary star spirals-in it might be on a slow rate, such that it survives until the envelope inflation phase. Therefore, in what follows we underestimate the allowed parameter space for the secondary star to accrete mass during the envelope inflation phase. Namely, in many cases more massive secondary stars than what condition (1) allows can also exist. These more massive stars can be massive neutron stars or even black holes.          

\begin{figure}
\begin{center}
\includegraphics[width=0.45\textwidth]{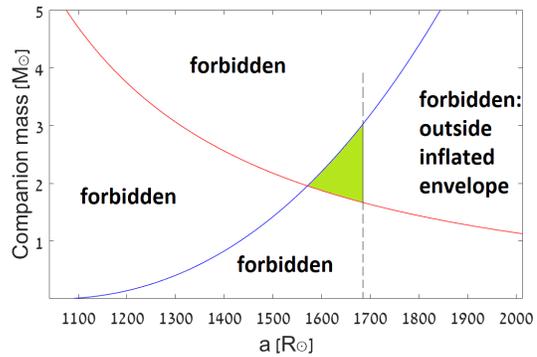}
\vspace*{-0.5cm}
\caption{  The interesting region in the plane of the mass of the secondary (companion) star versus the final orbital separation for the specific case we simulate here of a primary star with an initial mass of $15 M_\odot$, an initial metallicity of $Z=0.02$, and a wave power of $L_{\rm wave}=3.2 \times 10^{5} L_\odot$ that we inject into the envelope during the $2.3$ years before explosion. 
We plot in green the allowed region for the secondary star to survive before the envelope inflation phase and accrete mass during the inflation phase. 
The blue line is for the equality sign in equation (\ref{eq:RL1}), and the red line is for the equality sign in equation  (\ref{eq:Darwin}). The upper boundary of the green zone is the maximum radius the envelope reach during the inflation phase.   
}   
\label{fig:Darwin15}
\end{center}
\end{figure}

The effective radius of the Roche lobe of the primary star, as we take from 
\cite{Eggleton1983}, should obey 
\begin{equation}
R_{\rm L} = \frac{0.49q^{2/3}}
{0.6q^{2/3} + \ln \left( 1+q^{1/3} \right)} a > R_1(t=0), 
\label{eq:RL1}
\end{equation}
where $q=M_1/M_2$ in this case. 
In our single star evolutionary model $R_1=861 R_\odot$ just before the inflation phase, and $M_1=13.55 M_\odot$. Condition (\ref{eq:RL1}) implies that the system should be below the blue line in Fig. \ref{fig:Darwin15}. As we discussed above, in many cases a more massive secondary star can also survive until explosion. The upper limit on the secondary mass depends also on the response of the primary star to mass loss. Namely, if the primary expands to large radii as a result of mass loss it might engulf the secondary star. 
  
The moment of inertia of the star when we start the energy injection is $I_1=1.4 \times 10^6 M_\odot R^2_\odot$. Darwin stability, i.e., stability against a rapid spiraling-in process, reads 
\begin{eqnarray}
\begin{aligned}
M_2 > & \frac{3 I_1 (M_1 + M_2)}{M_1 a^2} = 4.2 
\left( \frac{I_1}{1.4 \times 10^6 M_\odot R^2_\odot} \right)
\\ & \times
\left( \frac{a}{1000 R_\odot} \right)^{-2} 
\frac{M_1 + M_2}{M_1} M_\odot
\end{aligned}
\label{eq:Darwin}
\end{eqnarray}
Condition (\ref{eq:Darwin}) implies that the system should be above the red line in Fig. \ref{fig:Darwin15}.

Another condition for the companion to accrete mass during he envelope inflation phase is that it will be within the inflated envelope during some portion of the envelope inflated phase. Namely, the orbital separation should be smaller than the maximum radius of the inflated envelope. In the present case this condition is $a<1690 R_\odot$, as marked by the vertical line in Fig. \ref{fig:Darwin15}.
The allowed region in the mass-orbital separation plane for a companion to accrete mass from the inflated envelope in our specific case is marked by green in Fig. \ref{fig:Darwin15}.  

The above case does not imply that the presence of an appropriate companion is very rare. In case of a massive enough companion, even if a RLOF does take place, the companion can remove enough mass from the primary envelope to cause it to shrink in a way that the primary gets back inside its Roche lobe. As well, in many cases the wave power can be larger and the expansion can be to a larger radius. 

\subsection{Accretion energy}
\label{sec:accretion}
  
 The allowed companion domain as we present in Fig. \ref{fig:Darwin15} fits a NS.
 Another possibility is that the secondary star is a low mass main sequence star if the primary star is the first star to become a CCSNe. A low mass main sequence star will at most lead to a small increase in luminosity. If it is limited to its Eddington luminosity, then the total accretion power will be $\simeq 5 \times 10^{4} L_\odot$, which is about the primary stellar luminosity. If all the mass as given by the accretion rate (see below) is accreted, then the total gravitational power for a low mass main sequence star is  $\simeq 3 \times 10^{6} L_\odot$. Even if all the energy is transferred to radiation (unlikely), this is still much below typical supernova luminosity. Over all, at best a small bump lasting several months might be observed in such cases, but only for objects that are in the nearby Universe. 
 
    
Because of the small influence of a main sequence secondary star, in section \ref{sec:companion} and below we consider a NS that orbits the primary star and accretes mass from the inflated envelope. 
 The typical relative velocity of the NS and the extended envelope is about the Keplerian velocity of the NS and the primary star   
 $v_{r} \simeq 44 [(M_1+M_2)/15M_\odot]^{1/2} (a/1500R_\odot)^{-1/2} \km \s^{-1}$. The Bondi-Hoyle-Lyttleton accretion rate is given by 
\begin{eqnarray}
\begin{aligned}
& \dot M_2 =  \pi \rho v_{r} \left(\frac{2GM_2}{v^2_{r}} \right)^2 
=  0.08 
\left( \frac{\rho}{10^{-9} \g \cm^{-3}} \right)
\\ & \times
\left( \frac{M_1+M_2}{15M_\odot} \right)^{-3/2}
\left( \frac{M_2}{1.4 M_\odot} \right)^{2}
\left( \frac{a}{1500 R_\odot} \right)^{3/2}
 M_\odot \yr^{-1}.
\end{aligned}
\label{eq:BHLaccretion}
\end{eqnarray}
We take the scaling of the density according to the results presented in Fig. \ref{fig:Density15}.

The accretion rate as given by equation (\ref{eq:BHLaccretion}) is about two orders of magnitude above the threshold for neutrino cooling to operate ($10^{-3} M_\odot \yr^{-1}$; \citealt{HouckChevalier1991}), and hence such an accretion rate is allowed despite being supper Eddington according to the usual definition. 
Consider that the NS accretes less than this value and power the inflated envelope with an energy of about $\eta_{\rm p} \simeq 0.1-0.01$ times the BHL accretion rate as given by equation (\ref{eq:BHLaccretion}). For an accretion phase that lasts for about one month, $t_{\rm acc} \simeq 0.1 \yr$, from the moment the inflated envelope reaches the secondary to explosion (see Fig. \ref{fig:Density15}), the total pre-explosion accretion energy liberated by the NS, as jets and radiation, is 
\begin{eqnarray}
\begin{aligned}
E_{\rm acc}  \simeq  10^{50}   
\left( \frac{\eta_{\rm p}}{0.1} \right)
\left( \frac{\dot M_2}{0.08 M_\odot \yr^{-1}} \right)
\left( \frac{t_{\rm acc}}{0.1 \yr} \right) 
\erg  .
\end{aligned}
\label{eq:BHLenergy}
\end{eqnarray}
 
We point out the following properties of the proposed interaction. 
\newline
(1) Most of the accretion energy is carried by neutrinos.  Most of the rest by jets. Typically, when compact objects launch jets, e.g., young stellar objects and black holes, about 10 per cent of the accreted mass is launched at a terminal velocity equals to the escape velocity. For that, the jets carry about 10 per cent of the accretion energy. As well, the accretion mass can be somewhat lower than that given by equation (\ref{eq:BHLaccretion}). This is the reason for the parameter of $\eta_p=0.01-0.1$ that we assume in equation (\ref{eq:BHLenergy}). Namely, it is possible that the energy carried by the jets is only $\approx 10^{49} \erg$ for the scaling we use here. 
Radiation carries negligible amount from near the NS as the inflow is optically thick. 
\newline
(2) As the jets collide with the inflated envelope and after they break-out from the envelope they interact with the previously blown wind. This interaction converts some kinetic energy to radiation and the process becomes visible. We cannot estimate the fraction of kinetic energy that is channeled to radiation as this is very sensitive to the distribution of the ambient gas with which the jets collide. 
\newline
(3) Since the inflated envelope grows to only about several times the initial radius of the star and there is not much mass available for accretion, the constraints on the companion are such that a NS companion will make much larger effects than what a main sequence stellar companion will make. 
\newline
(4) The interaction time is shorter than the orbital time. For the specific parameters that we are using here the orbital time at $a=1600 R_\odot$ is about $5 \yr$, while the phase of jets' launching at that radius lasts for less than half a year (right-most line in Fig. \ref{fig:Density15}). During that times the NS moves a distance of about $D \simeq 600 (t_{\rm acc}/0.3 \yr) R_\odot$. We speculate that the outcome of the interaction will be a highly distorted flow in that part of the inflated envelope where the NS resides, as the jets that the NS launches will expel the inflated envelope from that region only. We expect that the jets distort the structure of the envelope much as jets distort the envelope and winds of asymptotic giant branch stars that are progenitors of some planetary nebulae (e.g., as in the hydrodynamical numerical simulation of \cite{GarciaArredondoFrank2004} and \cite{Shiberetal2017}). If this interaction leads indeed to a distorted envelope, then the ejecta from the explosion that follows will interact with highly distorted inflated envelope and/or a shell. Such a geometry should be considered when fitting light curve of some CCSNe that are preceded by PEOs.

\subsection{Pre-explosion emission bump}
\label{sec:bump}
 
Let us elaborate on  the interaction of the jets with the inflated envelope  by conducting some simple estimates. 
Consider a NS at an orbital separation of $a= 1600 R_\odot$ that accretes mass from the inflated envelope for a time of $t_{\rm acc} \simeq 0.3 \yr$. During that time it moves a distance of about $D \simeq 600 R_\odot$ and accretes a mass of $M_{\rm acc} \simeq 0.026 M_\odot$ (by equation \ref{eq:BHLaccretion}). Assume the NS launches jets with a total energy of $E_{\rm jets} = 3 \times 10^{49} \erg$, i.e., $\eta_p = 0.01$ in equation (\ref{eq:BHLenergy}). This comes, for example, from a mass of $M_{\rm jets} = 3 \times 10^{-4} M_\odot$ in the two jets that is launched at a velocity of about $v_j \simeq 10^5 \km \s^{-1}$.  
Since the radius of the inflated envelope is $\simeq 1690 R_\odot$, the jets have a distance of $h \simeq 500 R_\odot$ to move perpendicular to the equatorial plane before they exist the inflated envelope. 

During  the interaction phase the two jets interact with an envelope mass of $M_{a} \approx 2 Dh^2 \tan \alpha_j \rho \simeq 0.03M_\odot$, where for the half opening angle of each jet we take $\alpha_j=30^\circ$, and we take the density of the inflated envelope $\rho(a=1600 R_\odot) \simeq 10^{-9} \g \cm^{-3}$ from Fig. \ref{fig:Density15}. Note that the location of the jets' axis constantly changes as the NS moves along its orbit, much like in simulations of a secondary star that launches jets in the wind or envelope of an asymptotic giant branch star (e.g. \citealt{GarciaArredondoFrank2004, Shiberetal2017}). 
Momentum conservation gives the approximate average velocity of the head of the jets $v_{\rm head} \approx v_j M_{\rm jet}/M_a \simeq 1000 \km \s^{-1}$. 
The jets break out from the envelope in a time of $t_{\rm break} \approx h/v_{\rm head} \simeq {\rm several \, day}$. Namely, the jets break out from the envelope, and the hot envelope gas in the outer interaction zones is exposed and radiates part of its thermal heat. This will lead to a transient event.  
 
 The breakout of the jets from the envelope will lead to X-ray emission as in a shock breakout from a wind in supernova explosion (e.g., \citealt{Svirskietal2012, Ohtanietal2018}). Unlike the case with a supernova where jets are launched from the center, here the jets move with the secondary star in its orbit. The duration of the X-ray emission lasts for about the duration of jets-launching phase. The physics is somewhat more complicated than that in supernova explosions and requires a separate study which is beyond the scope of the present paper.    
The jets shock the envelope gas they encounter and heat it. The heated gas cools by photon diffusion. The distance for the photons to diffuse out in the shortest way is $\Delta R \la 100 \AU$ for our case. The optical depth is $\tau \simeq  \Delta R \kappa \rho \approx 2 \times 10^3$, where $\kappa$ is the opacity which we take to be of electron scattering. The photon diffusion time is $t_\gamma \approx 3 \tau \Delta R/c \approx 15~$day. As the expansion time of the shocked inflated envelope gas is several days (the motion of the jets' head from origin to breakout $t_{\rm break}$), the gas suffers adiabatic losses. The fraction of the energy that is radiated away for our case is $E_{\rm rad} \approx t_{\rm break}/(t_{\rm break}+t_\gamma) \approx 0.2.$ The average luminosity over the $0.3 \yr$ is $L_{\rm rad} \approx 0.2 L_{\rm jets} \simeq 6 \times 10^{41} \erg \s^{-1}$. For an emitting area of 
$A \approx h^2 \simeq (500 R_\odot)^2$, and a black body emission,  the temperature is $5 \times 10^4 \K$. This is a strong UV source, but also strong in the visible, that lasts for several weeks. 

Over all, our proposed scenario leads to a several weeks long blue PEO that is a strong UV source and that is accompanied by X-ray emission.  
   
\section{Accretion from an expanding shell}
\label{sec:CSM}
\subsection{The scenario}
\label{subsec:scenario}

An accretion from an inflated envelope is not the only way to form a pre-explosion transient by a mass-accreting companion. Consider the model that \cite{Restetal2018} present for the fast-evolving luminous transient KSN 2015K. In that model a massive star at the end of its evolution lost a mass of $M_{\rm CSM} = 0.15M_\odot$ that at the time of explosion was a CSM shell with a radius of $R_{\rm CSM}=4\times 10^{14} \cm$ and a width of $\Delta R_{\rm CSM} = 1 \times 10^{14} \cm$. 
The supernova ejecta in that model has a mass of $M_{\rm ej}= 10 M_\odot$, a velocity of $v_{\rm ej} =8500 \km \s^{-1}$, and a kinetic energy of  
$E_{\rm ej} = 7 \times 10^{51}$. We note that neutrino-driven explosion cannot supply this energy (e.g. \citealt{Ebingeretal2018}), and another mechanism exploded the star. It is our view that this supernova was exploded by jets, as we think all CCSNe are (e.g. \citealt{SokerGilkis2017a}).  

Let us consider such a shell ejection in a case where a NS companion orbits the supernova progenitor. 
If such a CSM shell is ejected at the escape speed from the surface of the primary star $v_{\rm CSM}=(2GM_1/R_1)^{1/2}$, and the companion is not too close to the surface, then the relative velocity of the shell and the companion is $v_r \simeq v_{\rm CSM}$. The fraction of the mass that is accreted by the secondary star is $f_{\rm acc} \simeq \pi R^2_{\rm acc} /4 \pi a^2$, where $R_{\rm acc}$ is the accretion radius and $a$ is the orbital separation.  We find for the accreted mass 
\begin{eqnarray}
\begin{aligned}
M_{\rm acc-CSM} & \simeq \frac{1}{4} 
\left( \frac{M_2}{M_1} \right)^2
\left( \frac{R_1}{a} \right)^2 M_{\rm CSM}    = 4 \times 10^{-5} 
\\ & 
\left( \frac{M_2}{0.1 M_1} \right)^2
\left( \frac{a}{3 R_1} \right)^{-2} 
\left( \frac{M_{\rm CSM}}{0.15 M_\odot} \right) M_\odot .
\end{aligned}
\label{eq:Macccsm}
\end{eqnarray}
\cite{Restetal2018} consider two possibilities, that the star was a hydrogen poor progenitor with a radius of no more than several solar radii or that it was a giant. In the first case the shell velocity is $v_{\rm CSM} \simeq 1000 \km \s^{-1}$ and in the second it is $v_{\rm CSM} \simeq 100 \km \s^{-1}$. The accretion phase lasts for a time of 
\begin{equation}
t_{\rm acc-CSM} \simeq 0.3
\left( \frac{\Delta R_{\rm CSM}}{10^{14} \cm} \right)
\left( \frac{v_{\rm CSM}}{100 \km \s^{-1}} \right)^{-1} \yr .
\label{eq:tcsm}
\end{equation}
  
 Unlike the classical Bondi-Hoyle-Lyttleton accretion flow, here the flow deviates from a pure axi-symmetrical flow   because the density of the shell changes with radius and the secondary star has an orbital (transverse to the density gradient) velocity. As a result of that the accreted gas has a net angular momentum. Because of the very small radius of the NS, a very plausible outcome is the formation of an accretion disk. We assume below that an accretion disk forms and that it launches jets.  
  
\subsection{Stripped-envelope supernova progenitors}
\label{subsec:StrippedEnvelope}
For a hydrogen deficient progenitor of a small radius the duration will be 10 times shorter than that given by equation (\ref{eq:tcsm}), namely $t_{\rm acc-CSM} \simeq 0.03 \yr$.  For the parameters used in equation (\ref{eq:Macccsm}) the accretion rate onto a NS close to a small massive star would be about $M_{\rm acc-CSM}/t_{\rm acc-CSM} \approx 10^{-3} M_\odot \yr^{-1}$, hence neutrino cooling is efficient. Let us take the fraction of accretion energy on to the NS that goes into radiation, directly or first to kinetic energy and then radiation, to be $\eta_{\rm p} =0.1$. For the parameters used in equation (\ref{eq:Macccsm}) and for the potential well of a NS of $(10^5 \km \s^{-1})^2$, we find the emitted radiation from the NS and its jets interaction with the shell to be $E_{\rm rad-CSM} \simeq 10^{48} (\eta_p/0.1)  \erg$. The accretion phase lasts for about one to two weeks for the above parameters. However, the region is optically thick, and most of this energy will not escape from the system as radiation. 
  
In a case of a NS at a separation of $ \la \Delta R_{\rm CSM}$,  which is definitely the case here where $a < 100 R_\odot$, the typical optical depth of a spherical shell is $\tau > M_{\rm CSM} \kappa  /[4 \pi (\Delta R_{\rm CSM})^2] \simeq 10^3$, where we take electron scattering opacity and the CSM mass $M_{\rm CSM} = 0.15M_\odot$ as in section \ref{subsec:scenario}. The photon diffusion time out is 
\begin{equation}
t_\gamma \approx 3 \tau \Delta R_{\rm CSM}/c \approx 100~{\rm day}, 
\label{eq:tgamma}
\end{equation}
which is about ten times longer than the flow time $t_f \simeq \Delta R_{\rm CSM} /10^3 \km \s^{-1} \simeq 10$~day. 
As the expansion time is about 0.1 times the photon diffusion time, most of the energy that is released by the accretion process is doing work on the gas during the adiabatic expansion, and only about 10 per cent of the energy escapes as radiation. 
 
However, the energy carried by the jets can shape the shell. 
If about ten per cent of the accreted mass is ejected in jets at the escape speed from a NS of $10^5 \km \s^{-1}$, then the energy carried by the two jets is  
\begin{equation}
E_{\rm jets, NS} \simeq 0.1 M_{\rm acc-CSM} v^2_j/2 \approx {\rm few} \times 10^{47} \erg. 
\label{eq:Ejets}
\end{equation}
This energy is of the order of tens of per cent of the kinetic energy of the shell 
$E_{\rm CSM,1000} \simeq 1.5 \times 10^{48} \erg,$ 
for the parameters used here and a shell velocity of $v_{\rm CSM}=1000 \km \s^{-1}$. The jets will open two opposite small lobes (`ears'), along which the optical depth will be much lower. We might then have a transient event. Since the radius is small, it will be a blue event lasting for about days to few weeks, i.e., the flow time $t_f$ calculated above which is the time the shell pass the NS. If a fraction of $\simeq t_f/(t_f+t_\gamma) \approx 0.1$, of the kinetic energy of the jets is transferred to radiation, the typical luminosity of the event would be 
$L_{\rm rad-CSM} \approx 0.1 E_{\rm jets, NS}/t_f  \approx 10^{40} - 10^{41}\erg \s^{-1}$. 

\subsection{Giant progenitors}
\label{subsec:Giants}
\subsubsection{A NS companion}
\label{subsubsec:NS}
Let us then consider a massive giant star of a radius of $\simeq 2-4 \AU$ that ejects such a shell at a velocity of $\simeq 100 \km \s^{-1}$. The acccretion rate according to the parameters used here is $\simeq M_{\rm acc-CSM}/t_{\rm acc-CSM} \approx 10^{-4} M_\odot \yr^{-1}$. This does not allow an efficient neutrino cooling. But if we consider a somewhat denser shell and/or more massive one, and the jets can also carry energy out of the accretion flow (e.g., \citealt{Chamandyetal2018} for a main sequence star or a white dwarf accretor), then we might consider an accretion at this accretion rate on to a NS. The radiation diffusion time is $t_\gamma \approx 3~{\rm months}$ by equation (\ref{eq:tgamma}), and the expansion time is $t_f \simeq \Delta R_{\rm CSM} /100 \km \s^{-1} \approx 3$~months. The equality $t_\gamma \approx t_f$ implies that about half of the energy is radiated away.  
Over all, the outcome might be a transient event lasting several months with a typical luminosity of $L_{\rm rad-CSM} \approx 0.5 E_{\rm jets, NS}/t_f  \approx 10^{40} \erg \s^{-1}$, where the jets' energy is from equation (\ref{eq:Ejets}) and for the giant star $t_f \approx 3~$months. 
 
\subsubsection{A main sequence companion}
\label{subsubsec:MainS}
A comment is in place here.
Consider the formation of the shell in the model of \cite{Restetal2018} in the case of a giant. The shell was ejected $\simeq 4\times 10^{14} \cm / v_{\rm CSM} \simeq 1.3 \yr$ before explosion. Consider a case where there is no NS companion, but rather a main sequence companion instead of a NS. If the envelope is inflated as in our study in section \ref{sec:evolution}, but for a longer duration of several years, then a companion of $2 M_\odot$ at an orbital separation of $\simeq 1600 R_\odot$ could have accreted mass from the inflated envelope. We estimate that during the inflation time of about one year such a companion accretes a mass of $\simeq 0.1 M_\odot$ (see equation \ref{eq:BHLaccretion}). 
 
As the radius of a main sequence star is much larger than that of a NS, the accreted gas must have a larger specific angular momentum to form an accretion disk compared with the case of a NS companion. We assume that the density gradient in the extended envelope is large enough that the flow is highly asymmetrical and the gas has indeed a sufficient specific angular momentum to form an accretion disk. This can be the case at least when the extended envelope first reaches the main sequence companion, and the density gradient is much steeper than $\rho(r) \propto r^{-2}$. 
  
If about 10 per cent of the accreted mass of $\simeq 0.1 M_\odot$ is ejected in jets at the escape speed from the main sequence star, then the jets carry an energy of  
$E_{\rm jets} \approx {\rm several} \times 10^{46} \erg$. The kinetic energy of the shell is $E_{\rm CSM,100} \simeq 1.5 \times 10^{46} \erg$ for a shell velocity of $v_{\rm CSM}=100 \km \s^{-1}$. This implies that if the shell in the model studied by \cite{Restetal2018} was ejected by a giant, it could have been power by a main sequence binary companion (e.g., \citealt{McleySoker2014}). This speculative chain of processes requires a more detail study.  

\section{Summary}
\label{sec:summary}

 We examined some aspects of the binary scenario for PEOs. We inflated an envelope of a giant star about two years before core collapse by injecting energy to the envelope (section \ref{sec:evolution}). This energy injection mimics the effect of waves or magnetic activity from the core of the pre-collapse star (section \ref{sec:intro}). 
 
We found that for the general conditions that we have used, bright PEOs that occur months before explosion require a NS companion at an orbital separation of $\approx 1000-2000 R_\odot$. At larger separations the accretion rate is too low or does not exist, while for much shorter separations the orbit is unstable (Fig. \ref{fig:Darwin15}). The constraints on the companion mass (green area in Fig. \ref{fig:Darwin15}) allow for a NS, and possibly for main sequence stars above the blue line in Fig. \ref{fig:Darwin15}. Because of the low density of the inflated envelope (Fig. \ref{fig:Density15}), only mass accretion on to a NS can lead to a very energetic PEO. 
 
Based on preliminary simulations we estimate that calculations with a more consistent binary evolution will increase the parameter space allowed for the orbital separation and mass of the companion, particularly if the primary giant star is smaller in size. The primary star might be smaller if the companion removes large amount of mass from its envelope (e.g., the progenitor of SN 1987A). This is the subject of a future study. 
 
 We assume that the radiation of the event is energized by jets that the mass-accreting NS launches for a time period of about weeks to months (section \ref{sec:accretion}). The jets collide with the inflated envelope and after they exit the inflated envelope they interact with the older wind. This interaction converts kinetic energy to radiation (section \ref{sec:bump}). 
 
 As the jet-launching episode lasts for a time shorter than the orbital time in the cases we studied, the jets distort the inflated envelope and the CSM. The explosion will lose spherical symmetry and axial symmetry after the shock breaks out from the inflated envelope. The explosion of CCSNe that follow PEOs might lack an axial (cylindrical) symmetry. 
 
Main sequence star companions can also energize PEOs, but much weaker ones (section \ref{sec:accretion}). The typical energy of the jets can be $E_{\rm jets,MS} \approx 10^{46} \erg$ for the typical parameters that we use in this study rather than $E_{\rm jets,NS} \approx 10^{48}-10^{50} \erg$ for NS companions (section \ref{sec:accretion}). Such jets can shape the CSM of CCSNe. 
In section \ref{subsubsec:MainS} we discussed the possibility that a main sequence companion accretes mass from an inflated envelope and launches jets that influence the structure of an expanding shell. The shell in the model of \cite{Restetal2018} for the fast-evolving luminous transient KSN 2015K could have been shaped by such jets.

This study adds to the cases where NS can power PEOs. \cite{Gilkisetal2018} discuss the case where a NS enters the envelope itself and power a very strong outburst, termed a common envelope jets supernova (CEJSN) impostor. 

The main findings of this study is that companions, in particular NS companions, but not only NS, can energize the radiation of PEOs and shape the CSM to acquire highly asymmetrical structure lacking even axisymmetrical geometry.     

We thank an anonymous referee for very detailed and useful comments that substantially improved the manuscript. We acknowledge support from the Israel Science Foundation and a grant from the Asher Space Research Institute at the Technion.

\label{lastpage}
\end{document}